\documentclass[11pt,twoside]{article}
\usepackage{jltp}
\usepackage{wrapfig,graphicx}

\title{Strong vortex pinning in the low--temperature superconducting phase of 
(U$_{1-x}$Th$_{x}$)Be$_{13}$}

\author{Ana Celia Mota, Elisabeth Dumont\address{Laboratorium f{\"u}r Festk{\"o}rperphysik, 
ETH Z{\"u}rich, Switzerland}, and James~L. Smith\address{Superconductivity Technology Center, 
Los Alamos Nat. Lab., USA}}

\runninghead{Ana Celia Mota, Elisabeth Dumont, and James~L. 
Smith}{Strong vortex pinning in the low--temperature phase of (U$_{1-x}$Th$_{x}$)Be$_{13}$}
	   
\begin{document}

\begin{abstract}
We have found a sharp transition at $T_{c2} = 350$\,mK in the vortex 
creep rate of a single crystal of (U$_{1-x}$Th$_{x}$)Be$_{13}$ with 
$T_{c} = 523$\,mK ($x$ = 0.0275).  For $T \ll T_{c2}$, no creep of 
vortices is observed in a time scale of $10^5$\,s, while for $T_{c2} < 
T < T_{c}$, vortices creep at very high rates 
 ($30\,\%$ of decay from 
 a metastable configuration in the first $10^5$\,s at $T = 400$\,mK).  
The sharp transition occurs at the same temperature at which the 
second jump in the specific heat appears in these samples.  Similar 
low levels of creep rates have been reported by us in the low--$T$ 
superconducting phase of UPt$_{3}$\cite{andreas}.  

\end{abstract}

\maketitle


\section{INTRODUCTION}

\begin{figure}[h]
 \begin{center}\leavevmode
 \includegraphics{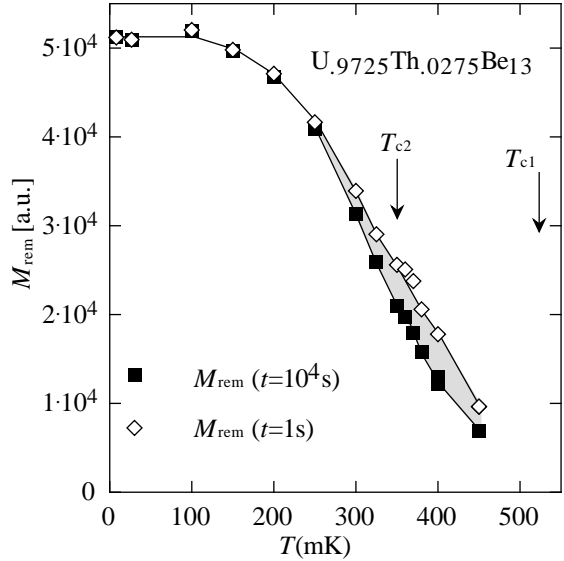}
 \caption{Remanent magnetization 
of U$_{.9725}$Th$_{.0275}$Be$_{13}$ at two different times as function of 
temperature. The lines are guide to the eyes.}
 \label{UTB_Mrem(T)}
 \end{center}
 \end{figure}

\begin{figure}[h]
 \begin{center}\leavevmode
 \includegraphics{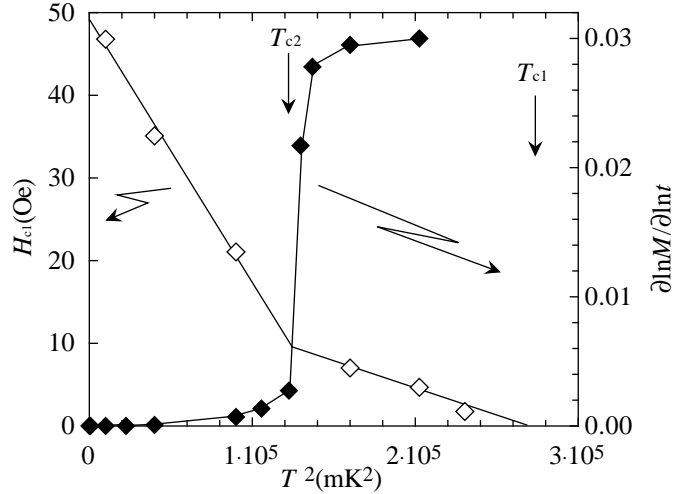}
 \caption{Initial creep rates (right scale, solid diamonds) and lower 
 critical field $H_{c1}$ (left scale, open diamonds) of 
 U$_{.9725}$Th$_{.0275}$Be$_{13}$ as function of $T^2$}
 \label{UTB_Hc1S(T)}
 \end{center}
 \end{figure}
 
 \begin{figure}[h]
 \begin{center}\leavevmode
  \includegraphics{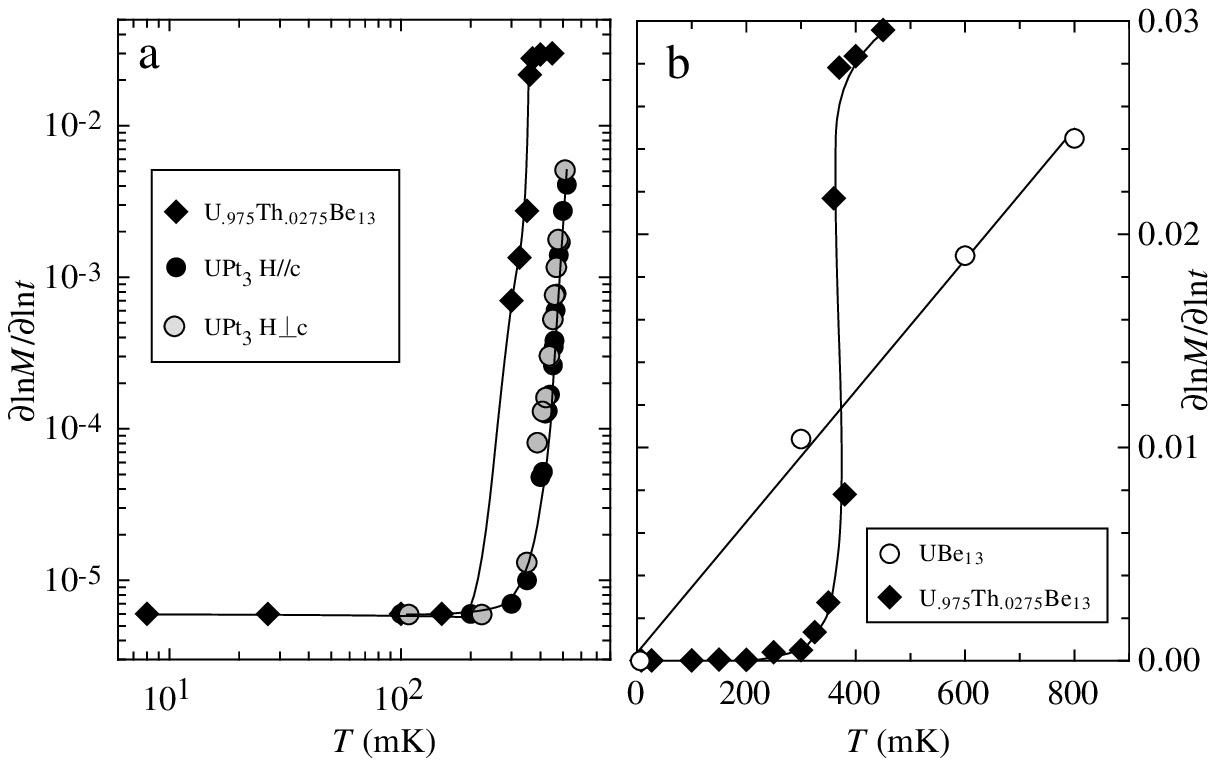}
 \caption{\,a. Transition of vortex creep rates in 
U$_{.9725}$Th$_{.0275}$Be$_{13}$ and in UPt$_{3}$ 
\,b. Creep rates in U$_{.9725}$Th$_{.0275}$Be$_{13}$ and in pure 
UBe$_{13}$}
 \label{UPt3UTB}
  \end{center}
  \end{figure}

%

Recent studies of vortex dynamics have uncovered a novel type of 
exceedingly strong vortex pinning not observed in any other hard 
type~II superconductor \cite{andreas}.  In the low--temperature, 
low--field superconducting phase of UPt$_{3}$, the so--called 
B--phase, metastable configurations of vortices do not creep towards 
equilibrium in a time scale of $10^4 - 10^5$\,s.  However, in the high 
temperature A--phase, vortex creep occurs with rates that increase 
rapidly as the temperature is increased.  The transition from one 
creep regime to the other is very sharp and more than two orders of 
magnitude in size.  The anomalous strong pinning detected only in the 
B-phase, indicates that this superconducting phase supports novel 
types of vortices and/or novel types of pinning structures.

The only other known example of a superconductor with more than one 
superconducting phase is thorium--doped UBe$_{13}$.  In the 
concentration range $0.019 \leq x \leq 0.045$, the heavy fermion 
(U$_{1-x}$Th$_{x}$)Be$_{13}$ shows an additional second order 
transition below the onset of superconductivity as first seen in 
specific heat measurements \cite{ott}.  Moreover, the lower critical 
field $H_{c1}$ shows a sudden break in slopes, indicating a clear 
increase in the superconducting condensation energy below the 
transition at $T_{c2}$ \cite{rauch}.  Muon spin relaxation data 
\cite{heff} reveal the existence of weak magnetic correlations in the 
low-T phase which are interpreted as evidence that this phase breaks 
time reversal symmetry.


Here we discuss the results of a recent investigation of vortex 
dynamics in a single crystal of (U$_{1-x}$Th$_{x}$)Be$_{13}$ with $x = 
0.0275$.  Similar results as the ones presented here were obtained 
with a second single crystal from a different batch.  We also compare 
the data on vortex creep with data on pure UBe$_{13}$ and UPt$_{3}$.

\section{EXPERIMENTAL ARRANGEMENT}
The sample investigated consisted of a single crystal 
(U$_{1-x}$Th$_{x}$)Be$_{13}$ with $x = 0.0275$ prepared at Los Alamos 
National Laboratory.  It was cut in the form of a parallelepiped $2.25 
\times 1.00 \times 0.88$\,mm$^3$ in size.  It has a transition 
temperature $T_{c}$ = 523 mK with a width $\Delta T_{c} = 67$\,mK taken 
with the $10-90 \%$\,criterion.

Vortex creep data was obtained from the relaxation of the remanent 
magnetization after cycling the zero-field cooled sample in high 
enough fields so that the critical state was established.  The 
relaxation measurements were done from 1\,s to $10^4 - 10^5$\,s in the 
temperature range 5\,mK$ \leq T \leq T_{c}$.  The experimental 
arrangement is described in reference~1.

\section{RESULTS AND DISCUSSION}

In Fig.\ \ref{UTB_Mrem(T)} we show values of the remanent magnetization 
of (U$_{1-x}$Th$_{x}$)Be$_{13}$ at two different times as function of 
temperature.  The shadowed area indicates the amount of flux that 
leaves the specimen from the initial time of our measurement ($t 
\approx 1$\,s) to $t = 10^4$\,s.  As can be clearly seen, deep in the 
low-temperature superconducting phase, no flux leaves the sample on a 
time scale of $10^4$\,s.  On the other hand, in the high temperature 
phase, considerably amount of vortices leave the specimen in the time 
indicated.  The initial creep rates from the data in Fig.\ \ref{UTB_Mrem(T)} are 
plotted in Fig.\ \ref{UTB_Hc1S(T)}.  In this figure we also show the 
measured values of the lower critical field $H_{c1}$ for the same 
crystal as function of $T^2$.  The sharp break of the $H_{c1}$ slope 
at $T = 350$\,mK occurs at the same temperature as the jump in the 
specific heat at $T_{c2}$ \cite{art}.  At this same temperature we 
observe a large transition from vortex creep rates $\partial \ln M/ 
\partial \ln t$ of the order of $3 \times 10^{-2}$ to values as small as 
 $\partial \ln M/ \partial \ln t \approx 10^{-5}$.  
This last figure 
reflects the limit of our sensitivity, determined mainly by the 
reproducibility of the background creep of the NbTi coil used to 
produce magnetic fields of the order of few hundred Oe.  This coil is 
directly attached to the walls of the Epoxy mixing chamber.

In Fig.\ \ref{UPt3UTB}\,a we compare the vortex transition in 
(U$_{1-x}$Th$_{x}$)Be$_{13}$ with similar transitions in UPt$_{3}$ 
\cite{andreas} on a double logarithmic scale.  Indeed, in both 
superconductors one detects "ideal pinning" in their low temperature 
phases.  However, although we do not observe creep on a scale of 
$10^5$\,s at the lowest temperatures, we detect some sort of 
"avalanche" or non logarithmic creep at times which become shorter and 
shorter as the temperature approaches $T_{c2}$.  The data in 
Fig.\ \ref{UTB_Hc1S(T)} and Fig.\ \ref{UPt3UTB} are based on the initial 
logarithmic slope calculated from the first couple of decades in time.

It is interesting to compare the vortex dynamics in pure UBe$_{13}$ 
with the dynamics of (U$_{1-x}$Th$_{x}$)Be$_{13}$ with $x=0.0275$ in 
view of some recent investigation by Kromer et al\cite{kromer}.  From 
thermal expansion measurements on samples of UBe$_{13}$ with different 
thorium concentrations, these authors concluded that the nature of the 
superconducting state in the critical concentration range ($0.019 \leq 
x \leq 0.045$) below $T_{c1}$ is not fundamentally different from 
that in the pure compound below $T_{c}$.

In Fig.\ \ref{UPt3UTB}\,b we show that the vortex creep rate in a 
single crystal of pure UBe$_{13}$ \cite{andreas} follows a well 
defined linear in $T$ dependence from $T=5$\,mK up to $T \approx T_{c} 
$ as expected from the Kim--Anderson theory of thermally activated 
creep.  No indication of anomalous strong pinning is detected in this 
material.  Indeed there is a fundamental difference between the 
low--$T$ superconducting phase of (U$_{1-x}$Th$_{x}$)Be$_{13}$ with 
$x=0.0275$ and pure UBe$_{13}$ concerning vortex dynamics.  

At this point the physical origin of "ideal pinning" is not known. 
However, by analogy with the superfluid phases of liquid $^3$He, where broken 
symmetries are manifested in new physical properties of the quantized 
vortex lines under rotation, one has to expect that in superconductors 
with nonscalar order parameters new types of vortices can also lead 
to unusual behavior.

In conclusion, we have found in UPt$_{3}$ and 
(U$_{1-x}$Th$_{x}$)Be$_{13}$ with $x=0.0275$ sharp transitions of 
about three orders of magnitude in the vortex creep rates at the 
temperature marking the boundary between their two low-field 
superconducting phases.  In both materials, deep in the low--$T$ phases  
no creep is observed on a scale of $10^5$\,s.  Theoretical input is needed to 
determine the physical origin of our observation.

\section*{ACKNOWLEDGMENTS}
We acknowledge valuable discussions with M. Sigrist and D. Agterberg. 
Part of this work was supported by the Swiss National Science 
Foundation.

\end{document}